\documentstyle[preprint,aps]{revtex}
\tighten

\begin{document}
\preprint{}


\title{The $J_1$ - $J_2$ model revisited : Phenomenology of CuGeO$_3$}
\author{V. N. Muthukumar$^1$, Claudius Gros$^1$, 
        Roser Valent\'\i$^1$,\\
        M. Weiden$^2$, C. Geibel$^2$, F. Steglich$^2$,
        P. Lemmens$^3$, M. Fischer$^3$, G. G\"untherodt$^3$
       }
\address{$^1$ Institut f\"ur Physik, Universit\"at Dortmund,
         44221 Dortmund, Germany\\
        }
\address{$^2$ FB Technische Physik, TH-Darmstadt, 
         Hochschulstr. 8, 64289 Darmstadt
        }
\address{$^2$ 3. Physikalisches Institut, RWTH Aachen,
         52056 Aachen, Germany
        }
\date{\today}
\maketitle
\begin{abstract}
We present a mean field solution of the antiferromagnetic Heisenberg
chain with nearest ($J_1$) and next to nearest neighbor ($J_2$) 
interactions. This
solution provides a way to estimate the effects of frustration. 
We calculate the temperature-dependent spin-wave velocity, $v_s(T)$
and discuss the possibility to determine the magnitude of
frustration $J_2/J_1$ present in quasi 1D compounds from
measurements of $v_s(T)$.
We compute the thermodynamic susceptibility at finite temperatures and
compare it with the observed susceptibility of the spin-Peierls compound
CuGeO$_3$. We also use the method to study the two-magnon Raman 
continuum observed in CuGeO$_3$ above the spin-Peierls transition.
\end{abstract}
%
%
\newpage

\section{Introduction}
The discovery of the spin-Peierls (SP) transition in the inorganic material
CuGeO$_3$ \cite{Hase_93} has led to an intense investigation
of the magnetic properties of this system. It is becoming increasingly
clear that this material may not be a prototype SP system such
as the quasi one-dimensional organic material TTFCuBDT \cite{Cross_79}.
For example, the
temperature dependence of the SP gap \cite{Martin_96} is unlike what 
is expected for conventional SP systems \cite{Cross_79}. 
A recent inelastic neutron scattering
experiment \cite{Lorenzo_96} reports observing a 
spin gap at temperatures above the SP transition temperature $T_{SP}$. 
So far, there has been no evidence for the presence of a phonon soft mode
\cite{Regnault_96}. 
X-ray scattering studies of the incommensurate phase of CuGeO$_3$ show the
existence of a soliton lattice with the width of the soliton being much
larger than that predicted by calculations \cite{Kiryukhin_96}. 
These results, taken together, call for a better understanding 
of the homogeneous state of CuGeO$_3$ above $T_{SP}$
which should eventually shed light on the nature of the SP transition itself.

The basic structure of CuGeO$_3$ consists of edge-sharing CuO$_6$ octahedra 
forming CuO$_4$ chains along the crystallographic \^{c} axis. The
dimerization of the $S = {1 \over 2}$ Cu ions below 14 K has been
determined by neutron-diffraction measurements \cite{Hirota_94}. From
inelastic neutron scattering (INS) measurements, Nishi {\it et al.} 
\cite{Nishi_94} estimated
the intrachain and interchain exchange parameters $J_c \approx$ 120
K, $J_b \approx$ 0.1$J_c$, suggesting that interchain effects may not
be negligible in this compound. On the other hand, INS
measurements above $T_{SP}$ \cite{Arai_96} clearly
show the two-spinon continuum which is characteristic of a one
dimensional $S = {1 \over 2}$ antiferromagnetic chain
\cite{Mueller_81}. The susceptibility of CuGeO$_3$ above 
$T_{SP}$ \cite{Hase_93} shows a broad maximum as expected for 
Heisenberg chains. From the temperature
at which this maximum is observed (56 K), one can estimate the value of
$J_c$ using the results of Bonner and Fisher \cite{Bonner_64}, 
yielding $J_c$ = 88 K. 
This discrepancy between the value of $J_c$ estimated
from static susceptibility measurements and INS led to the proposal 
\cite {Riera_95}, \cite{Castilla_95} that
a minimal model to describe the magnetic properties of this system 
above $T_{SP}$ is the so-called ``$J_1$ - $J_2$'' model. 
This model describes an
antiferromagnetic Heisenberg chain with nearest neighbor (n.n.) and
next-to-nearest neighbor (n.n.n.) exchange interactions. The model
Hamiltonian is written as
\begin{equation}
H = J \sum_i ({\bf S}_i \cdot {\bf S}_{i+1} + \alpha {\bf S}_i \cdot 
{\bf S}_{i+2})~~,
\label{ham}
\end{equation}
where $J \equiv J_c$ is the intrachain superexchange between neighboring
Cu ions along the \^{c} direction. The second term in the Hamiltonian
(\ref{ham}) is the exchange interaction between next to nearest
neighbor Cu ions. In CuGeO$_3$, the {n.n.n.} superexchange path is 
through Cu - O - O - Cu and is identical to that in the cuprate
superconductors. A detailed analysis of the structure of CuGeO$_3$ and
its relation to the magnetic interaction can be found in
\cite{Braden_96,Khomskii}.  

The model Hamiltonian (\ref{ham}) has been
studied by several authors \cite{several}. Though
the Hamiltonian is not exactly solvable for all values of $\alpha$, the
phase diagram is well understood qualitatively. For $0 < \alpha <
\alpha_{cr}$, the ground state remains gapless (as is the case when
$\alpha = 0$). The effect of $\alpha$ is to renormalize the spin wave
velocity in this regime. The value of $\alpha_{cr}$ has been estimated
to be 0.2411 \cite{Okamoto_93}. When $\alpha > \alpha_{cr}$, the spectrum
becomes gapped and for $\alpha = {1 \over 2}$, 
the Hamiltonian (\ref{ham}) is exactly solvable \cite{Majumdar_69}. 
As mentioned earlier, there has been a renewed interest in this model
in the context of CuGeO$_3$. Riera and Dobry \cite{Riera_95} 
as well as Castilla and co-workers \cite{Castilla_95}
computed the thermodynamic
susceptibility of the Hamiltonian (\ref{ham}) numerically and compared
it with the experimental values. Both groups found that the presence of
a non-vanishing $\alpha$ is needed to provide a consistent description
of both the INS and susceptibility results. 

If indeed the $J_1$ - $J_2$ model is an appropriate starting point to
embark on a study of CuGeO$_3$, it would be desirable to have some way
of calculating physical quantities, especially considering
the wealth of experimental data now available on CuGeO$_3$. 
This is the primary objective of this study where
we present some results obtained from a solitonic mean field
theory of the $J_1$~-~$J_2$ model. 
This method, which  is meaningful only for $\alpha < \alpha_{cr}$, 
provides a simple self-consistent way of evaluating
the effects of frustration. We use this solution to obtain the spinon
dispersion relation as a function of $\alpha$. We also calculate the ground
state energy and bulk susceptibility at finite temperatures as a
function of $\alpha$. The results for the bulk susceptibility 
are compared with the experimentally
observed values in CuGeO$_3$. We then examine the Raman continuum seen
experimentally above $T_{SP}$ and show how this arises as a
natural consequence of competing magnetic interactions.  We compute the Raman
intensity using the mean field solution and compare with experimental
results. The paper is organized as follows. In section II, we present
the mean field solution of the Hamiltonian (\ref{ham}) and use it to
compute the static susceptibility. In section III, we compute the two-magnon
Raman scattering intensity arising from competing magnetic interactions.
Section IV contains a brief summary of our results.
\section{A Mean Field Solution of the $J_1$ - $J_2$ Model}
In this section, we propose a mean field solution of the Hamiltonian
(\ref{ham}). This solution is based on a mapping introduced by
G\'{o}mez-Santos \cite{Gomez_90} between the spin-${1 \over 2}$ 
Heisenberg chain with {n.n.} interactions and a Hamiltonian 
describing the dynamics of antiferromagnetic domain walls. 
It has also been used by Weng and collaborators \cite{Weng_92} for the
one dimensional $t$ - $J$ model. We show below how the
mapping can be generalized to the spin Hamiltonian (\ref{ham})
with {n.n.n.} interactions as well. 
In this mapping, the local degrees of freedom
are given by the nature of the bond (ferromagnetic or antiferromagnetic)
between two interacting spins. We work using periodic boundary conditions and
choose the following convention: N\'eel ordering is characterized by an ``up
'' spin at the first site. Since, by definition,
the N\'eel ordered state does not have a ``kink" (henceforth, we shall use
the words ``kink'', ``soliton'' and ``domain wall'' interchangeably), 
this state is the vacuum state $\vert 0 \rangle$ of the solitons,
which we write symbolically as
$$
|0\rangle\, \equiv \, |\uparrow\downarrow
\uparrow\downarrow \uparrow\downarrow
\uparrow\downarrow \uparrow\downarrow
\uparrow\downarrow\dots
 \rangle~~.
$$
A state with a kink between sites $i$ and $i+1$ is defined to be 
a one-soliton state $d^{\dagger}_i \vert 0 \rangle$. For 
example,
$d^{\dagger}_4 \vert 0 \rangle$ defines the spin configuration
$$
d^{\dagger}_4 \vert 0 \rangle \, \equiv \, |\downarrow
\uparrow\downarrow \uparrow\uparrow\downarrow
\uparrow\downarrow \uparrow\downarrow
\uparrow\downarrow\dots
 \rangle~~.
$$

With these
definitions, it is easy to verify that the {n.n.} term in (\ref{ham}) is
mapped into
$$
{J \over 2} \sum_i \left\{
( d^{\dagger}_{i-1} d_{i+1} + d^{\dagger}_{i-1}
d^{\dagger}_{i+1} + {\rm h.c.} ) ( 1 - d^{\dagger}_i d_i) + 
(d^{\dagger}_i d_i - {1 \over 2})\right\}~~.
$$
Here, we have avoided the sign-problem by assuming the solitons to be 
hard-core bosons. On performing the Jordan-Wigner transformation
$$
d_i \rightarrow \exp {(i\pi \sum_{m<i}d^{\dagger}_md_m})\,d_i~~,
$$
one sees that the n.n. term written above preserves its form.
By looking at the action of the {n.n.n.} term in (\ref{ham}) on a pair of
interacting spins (or equivalently, a given bond), one can write down
the {n.n.n.} term in terms of the soliton operators in the same manner as
above.  Doing this and performing a Jordan-Wigner transformation, we 
obtain the $J_1$ - $J_2$ model in terms of fermionic soliton operators
as
\begin{eqnarray}
H & = & {J \over 2} \sum_i \left\{
( d^{\dagger}_{i+1} + d_{i+1} ) ( 1 - d^{\dagger}_i d_i) 
( d^{\dagger}_{i-1} - d_{i-1} ) +
        (d^{\dagger}_i d_i - {1 \over 2}) \right\}\nonumber \\
  & + & {\alpha J \over 2} \sum_i (d^{\dagger}_{i+2}+d_{i+2})
  (d^{\dagger}_{i+1}d_i + d^{\dagger}_id_{i+1})
  (d^{\dagger}_{i-1}-d_{i-1}) \label{sol} \\
  & + & {\alpha J \over 4} \sum_i (2d^{\dagger}_id_i-1)
  (2d^{\dagger}_{i+1}d_{i+1}-1)~~.
\nonumber
\end{eqnarray}
The above mapping may also be verified by using the definition of the
original spin operators in terms of the solitonic operators, viz.,
\begin{eqnarray*}
S^+_i & = & {1 \over 2}
(d^{\dagger}_{i-1}-d_{i-1})(d^{\dagger}_i+d_i)(1-2S^z_i)\\
S^-_i & = & {1 \over 2}
(d^{\dagger}_{i-1}-d_{i-1})(d^{\dagger}_i+d_i)(1+2S^z_i)\\
S^z_i & = & {1 \over 2}(-)^{i+1}\exp{ (i\pi\sum_{j<i}d^{\dagger}_jd_j})~~.
\end{eqnarray*}
Substituting the above expressions for the spin operators in
(\ref{ham}), we recover (\ref{sol}). It should be noted that
the maximum number of fermion operators occurring in (\ref{sol})
is four, both for the {n.n.} term and for the {n.n.n.} term. Longer 
ranged interactions such as ${\bf S}_i\cdot{\bf S}_{i+3}$ would,
on the other hand lead to terms containing products of
six or more fermion operators.

The Hamiltonian (\ref{sol}) is solved by treating the quartic terms in
mean field theory. We define the following averages that are determined
self consistently: $\bar{n} = \langle d^{\dagger}_i d_i \rangle$, 
$\Delta_1 = \langle d^{\dagger}_{i-1} d_{i+1} \rangle$ and 
$\Delta_2 = \langle d^{\dagger}_{i-1} d^{\dagger}_{i+1} \rangle$. 
In terms of these averages, the mean field Hamiltonian is given by
\begin{eqnarray*}
H_{MF} & = & \sum_i d^{\dagger}_id_i~\{{J \over 2}[1-2(\Delta_1+\Delta_2)]
             +\alpha J(2\bar{n}-1)\} \\
       & + & \sum_i(d^{\dagger}_{i-1}d_{i+1}+
	     d^{\dagger}_{i-1}d^{\dagger}_{i+1}
             + {\rm h.c.})~\{{J \over 2}(1-\bar{n})+\alpha J 
	     (\Delta_1+\Delta_2)\} \\
       & + & E_0~~,
\end{eqnarray*}
where N is the number of spins in the chain and $E_0$ is defined by
$$
E_0 = 
NJ(\Delta_1+\Delta_2) - {NJ \over 4} - NJ(1-\bar{n})(\Delta_1+\Delta_2)
-N\alpha J(\Delta_1+\Delta_2)^2 -N\alpha J(\bar{n}^2-{1 \over 4})~~.
$$
The mean field Hamiltonian can be simplified further by defining the
following two quantities
\begin{eqnarray*}
J_B & \equiv & {J \over 2}[1-2(\Delta_1+\Delta_2)] + \alpha J(2\bar{n}-1)
\\
J_A & \equiv & {J \over 2}(1-\bar{n}) + \alpha J(\Delta_1+\Delta_2)~~.
\end{eqnarray*}
Using these definitions and Fourier transforming, the mean field $J_1$ -
$J_2$ model is written as
\begin{equation}
H_{MF} = \sum_k (J_B + 2J_A\cos 2k) d^{\dagger}_kd_k + i\sum_k J_A\sin
2k (d^{\dagger}_kd^{\dagger}_{-k} - d_{-k}d_k) + E_0~~,
\label{mfham}
\end{equation}
where the lattice constant has been set to be unity. The mean field
Hamiltonian (\ref{mfham}) can now be solved by introducing the
Bogoliubov transformation
$$
d_k = u_k \alpha_k - iv_k \alpha^{\dagger}_{-k}~~,
$$
where
\begin{eqnarray*}
u_k & = & {1 \over \sqrt{2}} (1 + {\epsilon_k \over E_k})^{1 \over 2} \\
v_k & = & {1 \over \sqrt{2}} (1 - {\epsilon_k \over E_k})^{1 \over 2}
~~{\rm sgn} k~~,\\
\end{eqnarray*}
with
$\epsilon_k = J_B + 2J_A\cos 2k$, $\Delta_k = 2J_A\sin 2k$ and 
$E_k = \sqrt{\epsilon_k^2 + \Delta_k^2}$.
It is then easy to see that the Hamiltonian (\ref{mfham})
reduces to
$$
H_{MF} = \sum_k ~E_k \alpha_k^{\dagger}\alpha_k + {1 \over 2} \sum_k
(\epsilon_k-E_k) + E_0~~.
$$
On evaluating the mean field quantities and substituting them in the
definitions for $J_A$ and $J_B$ we get two mean field equations that
have to be solved for self consistency:
\begin{eqnarray}
J_B &=& {J \over 2}[1+{1 \over N}
         \sum_k {J_B\cos {2k}+2J_A \over E_k} (1-2n_k)] 
         - {\alpha J \over N}
       \sum_k {2J_A\cos {2k}+J_B \over E_k} (1-2n_k) 
       \label{mfeqn} \\
2J_A &=& {J \over 2}[1 +{1 \over N}
         \sum_k {2J_A\cos {2k}+J_B \over E_k} (1-2n_k)] 
         - {\alpha J \over N}
         \sum_k {J_B\cos {2k}+2J_A \over E_k} (1-2n_k)~~, 
\nonumber
\end{eqnarray}
where $n_k$ is the usual Fermi distribution function.
When $\alpha = 0$, the above equations reduce to the mean field
equations written down in \cite{Gomez_90} and \cite{Weng_92}. 
The solution in that case was obtained as $J_B = 2J_A$. By
inspection, we see that the solution of the mean field equations is
still given by $J_B = 2J_A \equiv \bar{J}$, where now, $\bar {J}$ is
modified by $\alpha$.  The dispersion relation is as before,
$E_k = 2\bar{J}\vert \cos {k} \vert$. On substituting the solution in
either of the above two equations, we determine $\bar {J}$ by the
equation
$$
\bar {J} = {J \over 2} \left[~1 + {1-2\alpha \over N} 
\sum_k \vert \cos {k}\vert~
{\rm tanh} (\beta \bar{J} \vert \cos k \vert)~\right]~~.
$$
\newpage
At zero temperature, the above expression gives us the 
dispersion relation, 
\begin{equation}
E_k = J [1 + {2(1-2\alpha) \over \pi}]~\vert \cos k \vert~~.
\label{disp}
\end{equation}
When $\alpha = 0$, (\ref{disp}) describes the dispersion
relation for spinons obtained by Faddeev and Takhtajan \cite{Faddeev_81}. 
As pointed
out in \cite{Weng_92}, the spinon velocity obtained from the mean field
theory of antiferromagnetic domain walls, namely
$(1+2/\pi)J\approx1.64J$ is quite close to the 
exact value ($\pi/2J\approx1.57J$). When $\alpha$ is
nonzero, our result shows that the spinon velocity is reduced because of
frustration arising from the {n.n.n.} interaction. This is to be expected
from physical grounds. On comparing the spinon velocity in (\ref{disp}) 
with the spinon velocity for $\alpha = 0$, we see that the effect of the
n.n.n. term is to reduce the spinon velocity as 
$v_s(\alpha) = v_s(0)[1 - 4\alpha/(\pi+2)]$. 
The ground state energy at $T =
0$ is given by $E_0 = E_0(\alpha = 0) + 
2\alpha NJ/\pi^2$, where $E_0(\alpha = 0)/NJ = 
-\left(1/\pi+1/\pi^2\right)\approx-0.420$ 
is the ground state energy of the {n.n.} Heisenberg
chain within this approach, which compares well with
the exact result $1/4-\ln 2 \approx -0.443$ \cite{Lieb_Wu}.

The mean field solution described above provides a simple way to
calculate physical quantities in the $J_1$ - $J_2$ model. In particular,
one can see how the presence of the {n.n.n.} term in the Hamiltonian
(\ref{ham}) alters the susceptibility at finite temperatures. Before we
proceed to show this, we discuss the limitations of the mean field
theory. First, we point out that though the mean field {\it solution} is
valid for any value of $\alpha$, the theory itself is meaningful only
in the gapless regime, {\it i.e.}, when $\alpha < \alpha_{cr}$ where the
spinons are deconfined. This mean field theory is not suited for studying
the transition between the gapless and gapped phases of the $J_1$ -
$J_2$ model. We reemphasize that our objective is not 
to undertake such a study (indeed, far more powerful techniques are
available and have been used to study this problem) but to provide a
simple self consistent prescription to calculate physical quantities in
the gapless phase of the $J_1$ - $J_2$ model. Next, we point out that
the mean field solution is plagued by its reluctance to move away from
the universality class of the $XY$ model. This can be seen by writing
down the (mean field) ground state wave function in terms of the soliton
operators $d_k$ as 
${\textstyle {\vert \Psi \rangle = \prod_k( u_k + v_k d^{\dagger}_k
d^{\dagger}_{-k} ) \vert 0 \rangle}}$. 
Whereas in the $XY$ model, the
Hamiltonian can be solved exactly to determine $u_k$ and $v_k$, the mean
field solution of the Heisenberg as well as the $J_1$~-~$J_2$ model give
the same $u_k$ and $v_k$. In this sense, this theory has the same
difficulties as the mean field theory of Bulaevski\v{i}
\cite{Bulaevskii_63} though the nature of the quasiparticles in the two
theories are entirely different.
Finally, let us consider the reduction in the spinon velocity as a
function of the {n.n.n.} coupling $\alpha$. 
From (\ref{disp}), we see that the velocity $v_s(\alpha)
\approx v_s(0) (1-0.8\alpha)$. This result can be checked numerically.
In a recent numerical study, Fledderjohann and Gros \cite{Spock_96} 
found that the spinon velocity in the $J_1$ - $J_2$ model can be fit 
to the relation $v_s(\alpha) = v_s(0) (1-1.12 \alpha)$. Thus we see that the
results from mean field theory {\em underestimate} the effects of
frustration.

We are now ready to consider the static susceptibility at finite
temperatures. In particular, we are interested in seeing how the
presence of a {n.n.n.} interaction alters the susceptibility of an
antiferromagnetic Heisenberg chain. Now, it is well known that the
susceptibility of the Heisenberg chain is constant at zero temperature
and increases to a broad maximum at $T = 0.641 J$ \cite{Bonner_64}.
The increase in susceptibility is due to the gapless spinon excitations.
As we shall see, the {n.n.n.} term causes a suppression in the maximum
value of this susceptibility. The reason for this is that the spinon
velocity now has a different temperature dependence with a nonzero
$\alpha$. To illustrate this, we show in Fig.\ \ref{vel}, the spinon
velocity as a function of temperature. The
long-dashed line shows the results for the case $\alpha = 0$. The solid
line shows the spinon velocity as a function of temperature for $\alpha
= 0.24$. For the latter case, we have, for purposes of illustration,
chosen $J$ such that the velocity at
$T = 0$, given by (\ref{disp}) is the same as that for $\alpha = 0$.
It should be noted that while the low temperature velocities for different
$\alpha$ are identical, the results begin to differ as temperature
increases. In particular, they are perceptibly different at $T \approx
0.6 J$. Therefore, we should expect a difference in the finite
temperature susceptibilities as a function of $\alpha$ at these
temperatures.

We have calculated the static susceptibility at 
finite temperatures from the mean field solution.
At this point, it is crucial to
realize that the results from the mean field theory violate
spin-rotational invariance of the underlying Hamiltonian,
and that the physically relevant susceptibility would be given 
by the average
$$
\chi\, =\, {1\over3}(\chi_{xx}+\chi_{yy}+\chi_{zz})~~,
$$
where 
$\chi_{xx},\ \chi_{yy},\ \chi_{zz}$ are the
susceptibilities for the applied magnetic fields
in $x,\ y$ and $z$ directions respectively. Next we note that
our mean field theory becomes exact for the
$XY$ model for which the susceptibilities
$\chi_{zz}^{XY}=1/\pi\approx 0.318$ and
$\chi_{xx}^{XY}=\chi_{yy}^{XY}\approx0.075\approx 
 \chi_{zz}^{XY}/4$
\cite{Mueller_84}, in dimensionless units and
at zero temperature. We therefore have for the
$XY$ model, the relation
\begin{equation} 
\chi\, \approx\, \chi^{XY}_{zz}/2~~.
\label{factor_2}
\end{equation}
We have applied this relation to our mean-field
solution of the Heisenberg model.

A straightforward calculation shows that the
susceptibility at zero temperature is given
by
$$
{\chi \over g^2\mu_B^2 N} = {1 \over 2\pi v_s}~,
$$
where $g$ is the gyromagnetic ratio, $\mu_B$ is the electron Bohr
magneton and $v_s$, as before, is the spinon velocity. 
The factor $2$ in the denominator comes
from the average over the directions of the
applied magnetic field as given by Eq.\ (\ref{factor_2}).

As a quick check of our
results, let us consider the case $\alpha = 0$ and compare
with the results of Bonner and Fisher \cite{Bonner_64}. At $T = 0$, from
(\ref{disp}) and the above expression for susceptibility, we get
$\chi(T=0) J/(g^2\mu_B^2 N) \approx 0.106$, 
which compares very well with
the Bethe-Ansatz results $1/\pi^2\approx0.101$ \cite{Bonner_64}.
At finite temperatures, we
solved the mean field equations (\ref{mfeqn}) numerically and 
used it to determine the susceptibility. We found that
the temperature at which the susceptibility attains its maximum value is
given by
$k_BT_{{\rm max}}/J = 0.61$, 
while the corresponding value quoted in \cite{Bonner_64} is 0.641.
These comparisons encourage us to proceed to the case $\alpha \neq 0$.

Our results for the case $\alpha \neq 0$ are summarized in
Fig.\ \ref{susc_1D} where they are compared with experimentally obtained
values from a polycrystalline sample of
CuGeO$_3$ \cite{Weiden_96}. We have chosen $g
= 2.0$ in all our calculations.
The long-dashed line shows the results for $J =
88$ K, $\alpha = 0$ that are clearly in disagreement with experiment. The
choice of $J = 88$ K \cite{Hase_93} stems from the observed temperature 
(56 K) at which the
susceptibility attains its maximal value and its relation to J from the
results of \cite{Bonner_64} or equivalently, our results. 
The dotted line shows the results we obtained from
choosing $J = 115$ K, $ \alpha = 0$. This choice of $J$ yields the 
observed value of
spinon velocity from INS \cite{Nishi_94} without including the effects
of frustration. It is clear that there still remains a discrepancy
between the observed and calculated values at intermediate temperatures.
Our results for the susceptibility with a non vanishing $\alpha$ are
given by the solid line. Here, we have chosen $J = 142$ K and 
$\alpha = 0.24$. This choice of parameters when substituted in
(\ref{disp}) gives the value of the spinon velocity observed in INS.
(Note that the corresponding value of $J$ 
determined numerically in \cite{Castilla_95} 
is $J = 150$ K.)
At low temperatures, results with and without
$\alpha$ are identical (see Fig.\ \ref{vel} and our discussion of those
results). However we see that at as temperature increases, a non
vanishing $\alpha$ does tend to suppress the susceptibility. We see that
there is still a discrepancy between theory and experiment which is of
the order of 10\%. We believe this discrepancy is because the mean field
theory underestimates the effects of frustration. 
We also considered the
possibility that the discrepancy between theory and experiment may well 
be due to the effect of interchain coupling. As mentioned earlier, the
interchain superexchange $J_b \approx 0.1 J_c$ and this could 
play an important role. A qualitative argument against this possibility
is that at temperatures where the discrepancy is maximal ($T > 0.5
T_{SP}$), the magnetic
correlation lengths should be small. To substantiate this argument, we 
did a simple calculation using
renormalized spin wave theory with the appropriate values of $J_b$ and
$J_c$. Were it to order antiferromagnetically, the observed values of the
superexchange interactions suggest that the ``N\'{e}el temperature"
of CuGeO$_3$ would be $\sim 10$ K which is lower than $T_{SP}$.
At temperatures 
above $T_{SP}$, the effect of (short range) antiferromagnetic ordering 
can be easily estimated by spin wave theory. A standard calculation
using the ``decoupling approximation'' \cite{Balucani_76} gives us the
susceptibility $\chi$ in terms of the integral equation
$$
{1 \over \pi^2} \int_0^{\pi} \int_0^{\pi} dz~dy {1 \over 2(\gamma_o -
\gamma_k) - \chi^{-1}} = {1 \over 4k_BT}~,
$$
where $\gamma_o \equiv J_c+J_b$ and $\gamma_k \equiv J_c \cos  z + J_b
\cos y$. We have ignored $J_a \approx -0.01 J_c$ and set $g\mu_B$ to be
unity for convenience.  One integration in the {l.h.s.} of the above
expression can be done trivially and we obtain
$$
\int_0^{\pi} dz {1 \over \sqrt{(\bar{\chi} + \cos z)^2 - 0.01}} = 
{\pi \over 2}\left({J \over k_BT}\right)~~,
$$
where $\bar{\chi} \equiv (2J\chi)^{-1} - 1.1$ and $J_c \equiv J = 10J_b$.
Solving the above equation numerically with $J = 115$ K, 
we obtain the susceptibility as shown in Fig.\ \ref{susc_2D}. 
It should be remembered that the theoretical results are valid only at
temperatures well above $T_{SP}$ where the effects of phonons as well as 
effects due to competing antiferromagnetic and SP instabilities can be ignored. 
The results in Fig.\ \ref{susc_1D} and Fig.\ \ref{susc_2D} show that
a one dimensional model with frustration is indeed a better starting
point to study CuGeO$_3$ rather than an anisotropic two dimensional model.
This is in agreement with a number of experimental observations such as
(i) a non-zero nuclear spin-lattice relaxation rate seen in a Cu NQR
study \cite{Kikuchi_94}, (ii) the two-spinon continuum observed in INS
above $T_{SP}$ \cite{Lorenzo_96,Arai_96}, and (iii) a broad
continuum of magnetic excitations seen in two-magnon Raman scattering above
$T_{SP}$ \cite{Raman}, which will be discussed in the next section.
All these experiments show characteristic features of a one dimensional
Heisenberg antiferromagnetic chain. 

Since our mean field solution underestimates the effects of frustration,
we are unable to determine the exact ratio of $J_2/J_1$ in CuGeO$_3$. As
we mentioned in the introduction, the value of $J_2/J_1$ was estimated
to be 0.36 \cite{Riera_95} and 0.24 \cite{Castilla_95} by two
independent studies. Since these two values lead to two different fixed
points (gapped and ungapped spin excitation spectra respectively),
careful experimental investigation of spin dynamics above $T_{SP}$ may
provide further clues. Recently, Kuroe {\it et al.} \cite{Kuroe_96}
have claimed that $J_2/J_1 = 0.35 \pm 0.05$ leads to a good fit of the
magnetic specific heat as determined by quasi elastic Raman scattering.
It should be noted, however, that this choice does not fit the 
observed magnetic susceptibility. 
CuGeO$_3$ is 0.35 $\pm$ 0.05.

Before we conclude this section, we discuss how the temperature
dependence of the spinon velocity may be used to determine the presence
of competing magnetic interactions in quasi 1D materials. As shown in
Fig.\ \ref{vel}, a non zero value of $J_2/J_1$ leads to a difference in
the temperature dependence of $v_s(T)$. This feature may be detected
experimentally. As a possibility, we suggest INS experiments in the
presence of a magnetic field. In the presence of a magnetic field, it is
known that the spinon dispersion $E_k$ goes to zero at $k$ given by
$\cos k = -2H/v_s(T)$ \cite{Ishimura_77}. This is a feature of low
dimensional spin excitations and is seen in the $XY$ model as well. 
This behavior should be contrasted with spin wave (magnon) excitations,
where no such node appears in the magnon dispersion as a function
of the magnetic field. An examination of the temperature dependence 
of this node with moderately high fields $H/J \sim 0.2$ 
should indicate the presence of frustration as shown in Fig.\ \ref{vel}.
\section{Two-Magnon Raman continuum in CuGeO$_3$}

In this section, we consider the Raman spectra in the homogeneous phase
of CuGeO$_3$ observed in two-magnon scattering experiments \cite{Raman}. 
A partial account of the results in this section has been published
elsewhere \cite{rc}. For completeness, we first give a brief summary of
the pertinent experimental results.

Scattering of light from magnetic (as opposed to phonon)
excitations in CuGeO$_3$ 
is strongest in the $(zz)$ geometry {\it viz.}, when both incident
and scattered light are polarized along the crystallographic \^c
direction. This is consistent with the fact that exchange interactions
are strongest along this direction.  
The Raman spectra above $T_{SP}$ are rather
featureless and the spectral weight is distributed over a wide range of
energies (100-300 cm$^{-1}$). A broad maximum is observed around 250
cm$^{-1}$.  This should be contrasted with the Raman
spectra observed in antiferromagnetically ordered compounds such as
the rutile or perovskite halides. In these compounds,
one typically observes a well defined peak at characteristic magnon
energies which is broadened by
magnon-magnon interactions \cite{Cottam_86}. This comparison
immediately suggests light scattering in CuGeO$_{3}$ from a continuum of
excitations rather than single-particle excitations (magnons). 

The magnetic Raman scattering in CuGeO$_3$ can be studied
using the scheme of Fleury and Loudon \cite{Fleury_68}. The basic idea 
is that in an exchange-coupled system, an incident photon can create
particle-hole excitations accompanied by a pair of spin flips that
propagate as magnons. In this scheme the Raman operator $H_R$ has
essentially the same form as the Heisenberg exchange interaction,
{i.e.} $H_R\sim\sum_i{\bf S}_i\cdot{\bf S}_{i+1}$.
In one dimensional chains
with {n.n.} exchange interaction, 
we are faced with an intriguing situation where
the Raman operator for scattering in the $(zz)$ geometry is 
proportional to the Hamiltonian itself. Thus, there should
be no Raman scattering from a {n.n.} Heisenberg spin chain, a
fact which has not been appreciated previously \cite{Raman}.
However, there are two
ways in which Raman scattering can occur: (i) interchain couplings
and (ii) exchange interactions that are extended in space, {i.e.}
{n.n.n.} interactions. Let us first consider the former possibility. 
The role of the interchain couplings is to establish
short range antiferromagnetic order.
In such a case, incident light is scattered off paramagnon excitations.
However as we mentioned earlier, the featureless spectra observed in
CuGeO$_3$ seem to indicate scattering from a continuum of excitations.
This is basically a one-dimensional feature. The second possibility is 
therefore a more viable explanation of the experimental results.
In view of the results discussed in the previous section, it is also a
more natural choice. Thus, we work with a Hamiltonian which has both
{n.n.} and {n.n.n.} interactions. This implies that the Raman operator
will also have both n.n. and n.n.n. terms. For convenience, we subtract
from the Raman operator, a part which
commutes with the basic Hamiltonian of CuGeO$_3$ (\ref{ham}), we can
write down a Raman operator of the form
$$
H_R = A  \sum_i {\bf S}_i \cdot{\bf  S}_{i+2}~~.
$$
The Raman intensity is then evaluated by
considering the appropriate spectral function. 

We now use the mean field theory described in the previous section to
compute the Raman intensity. The Raman operator $R$ can be written in
terms of the soliton operators and transformed into quasiparticle or
spinon operators just as we did in the previous section. We find that there 
are three basic physical processes caused by $R$: (i) spinon-spinon
scattering, (ii) two-spinon creation/annihilation and (iii) four-spinon 
creation/annihilation. Of these, (ii) corresponds to one-magnon
excitations. On using the mean field solution, one can show that the
contribution to light scattering from these excitations vanishes, which
is just the statement that one-magnon excitations with $k=0$ have zero
spectral weight. Two-magnon scattering is given by (iii). We find that
the matrix element for two-magnon (four-spinon) creation 
(Stokes component of the scattering intensity) is given by
\begin{eqnarray*}
{1 \over N} \sum_{kk^{\prime}qq^{\prime}} &
\delta(k+k^{\prime}+q+q^{\prime}) & \alpha^{\dagger}_k~
\alpha^{\dagger}_{k^{\prime}}~ \alpha^{\dagger}_q~ 
\alpha^{\dagger}_{q^{\prime}} \\
& [ {1 \over 2} \exp\{{-i(2k+k^{\prime}-q^{\prime}})\} &
\{(u_ku_{k^{\prime}}+iv_ku_{k^{\prime}})
(v_qv_{q^{\prime}}+iv_qu_{q^{\prime}}) \\
& & 
-(u_kv_{k^{\prime}}+iv_kv_{k^{\prime}})
(u_qv_{q^{\prime}}+iu_qu_{q^{\prime}}) \} \\
& &
- \exp\{-i(q+q^{\prime})\}\, u_kv_{k^{\prime}}u_qv_{q^{\prime}}~]~~,
\end{eqnarray*}
where the summation is over the range of spinon momenta and the
quantities $u_k$ and $v_k$ are as defined in the previous section.
The matrix
element can be understood as arising from a decomposition of two magnon
excitations into four spinon excitations with total momentum zero. Given
this matrix element, we can calculate the Raman intensity. The mean
field solution thus provides a way of incorporating matrix element
effects arising from the Raman operator instead of postulating {\it ad
hoc} filtering functions that mimic matrix element effects.
As noted elsewhere \cite{rc}, such effects
could be very important and ignoring them 
often leads to erroneous conclusions.

\begin{figure}
\caption{The two-magnon Raman continuum seen in experiment 
(filled squares, \protect\cite{rc}, see also
\protect\cite{Raman})
at $T = 20$ K compared with results obtained 
from mean field theory (solid line). 
The theoretical results were obtained with $J = 142$ K and $\alpha = 0.24$.
\label{raman}}
\end{figure}
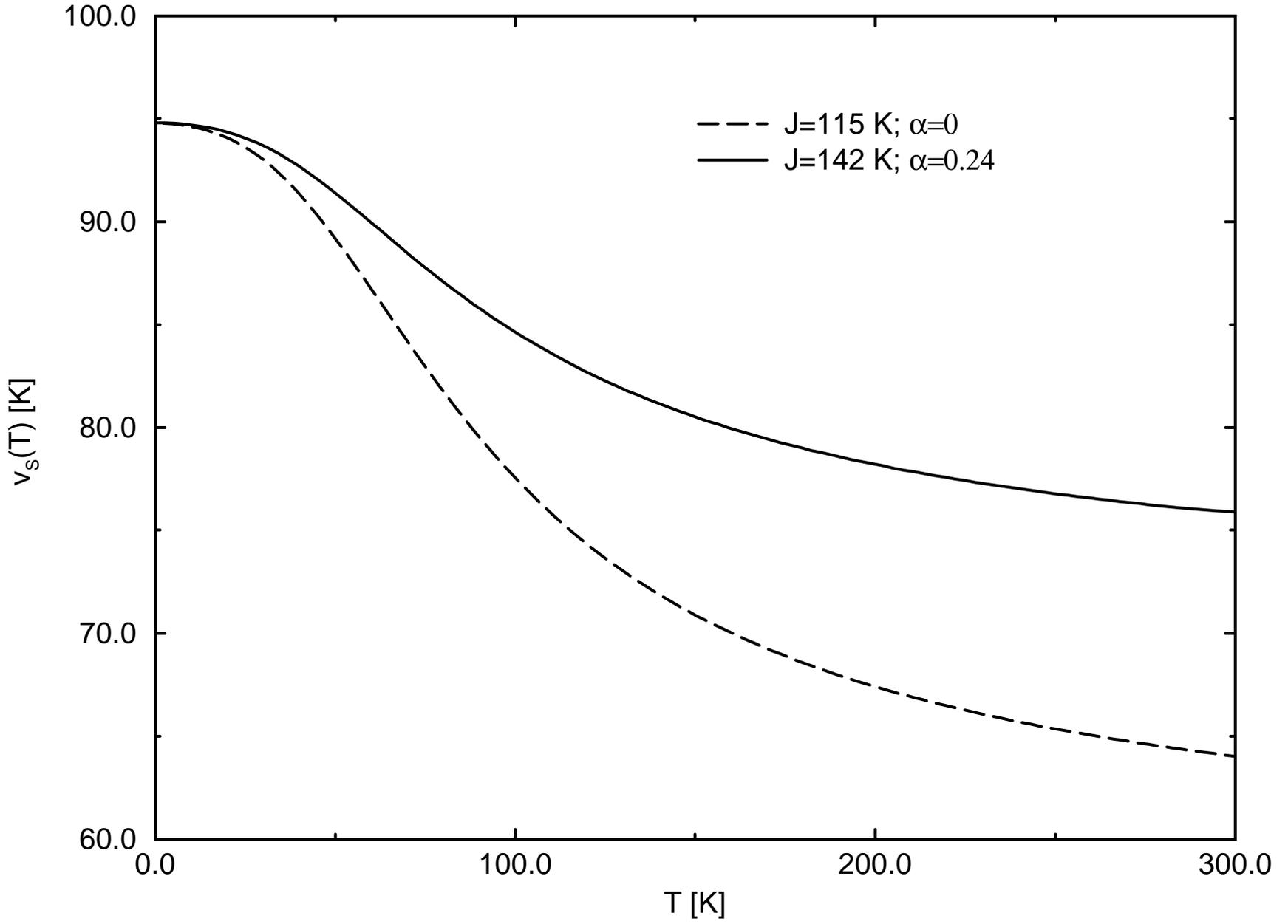
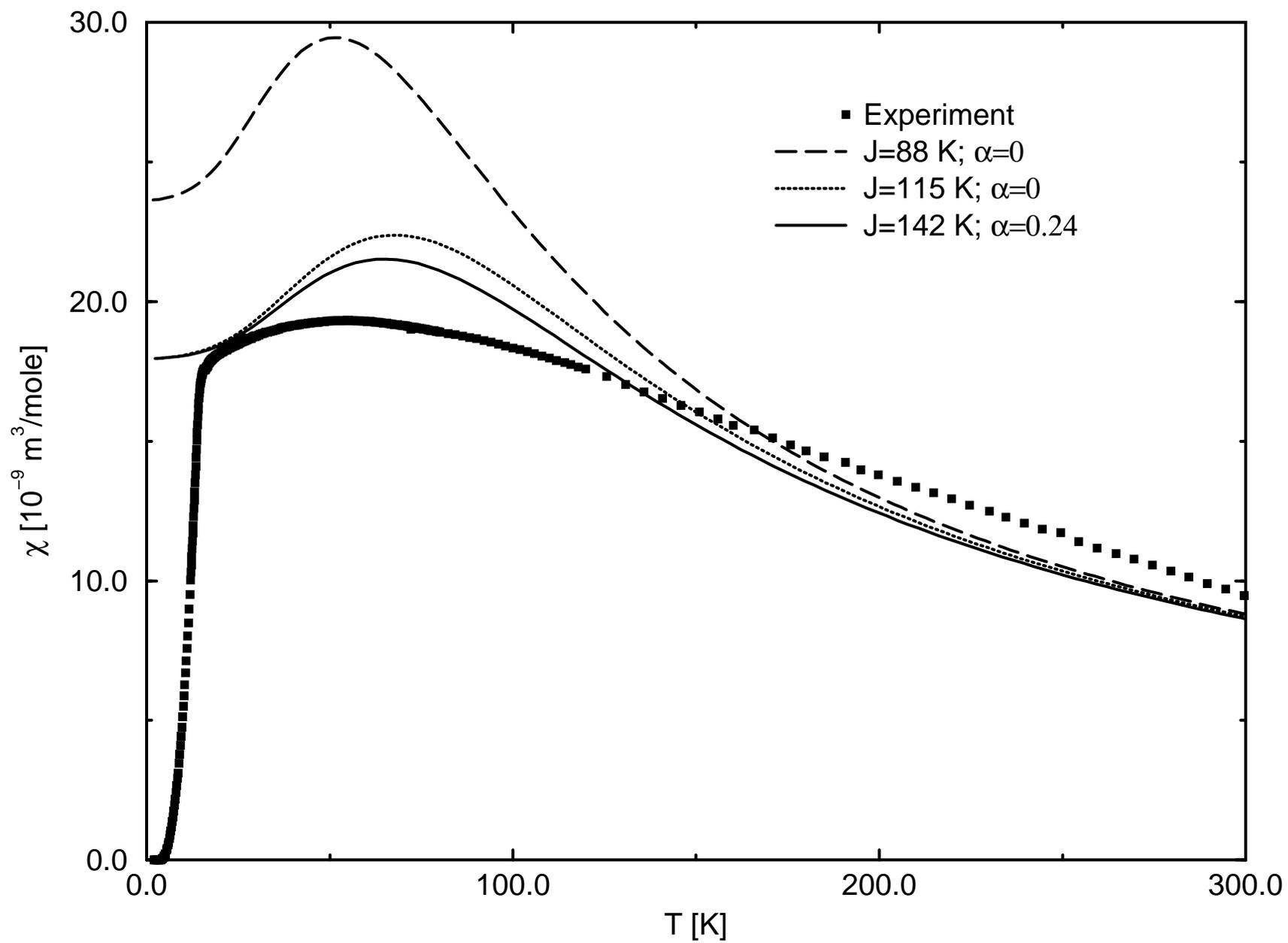
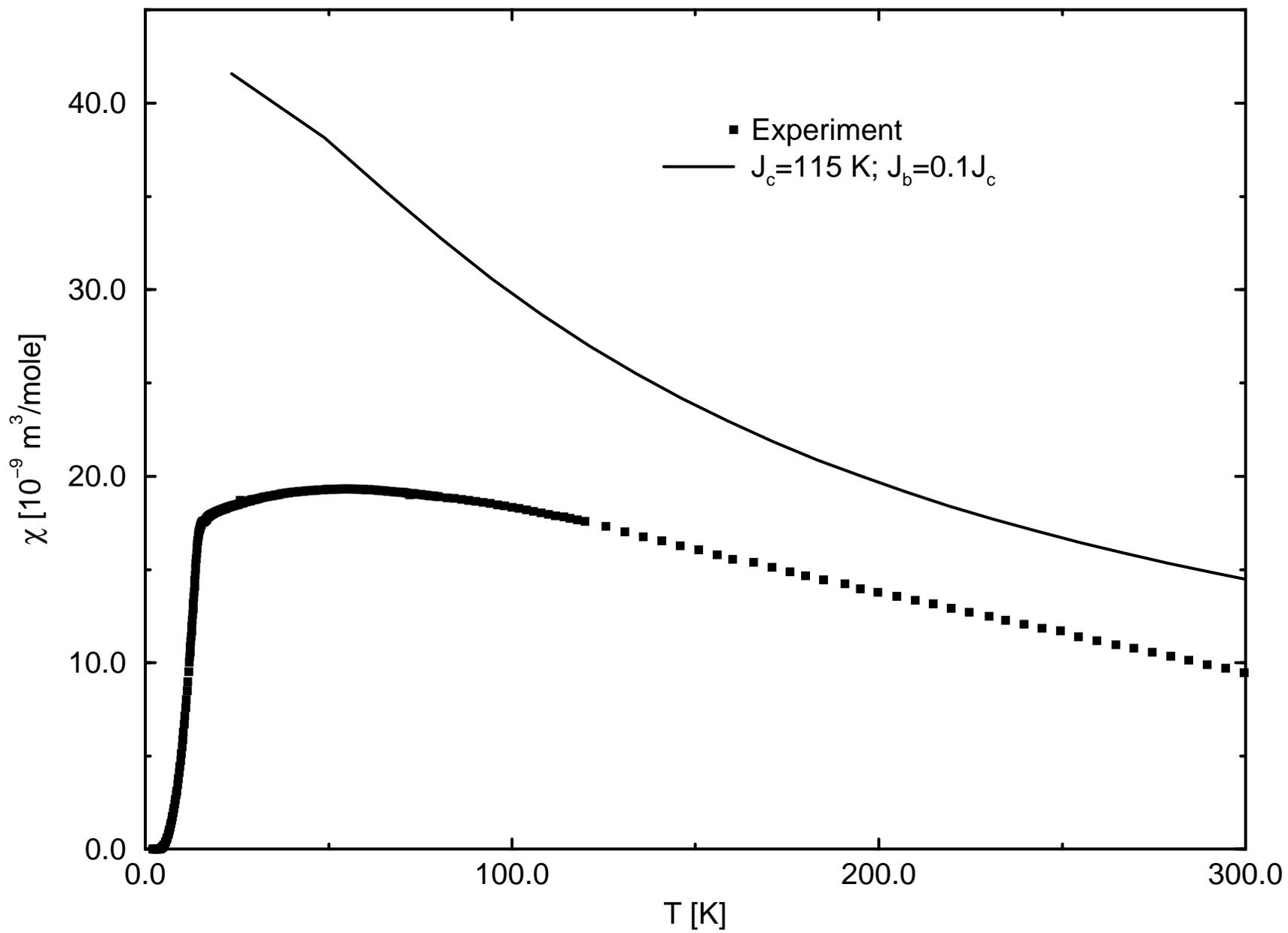
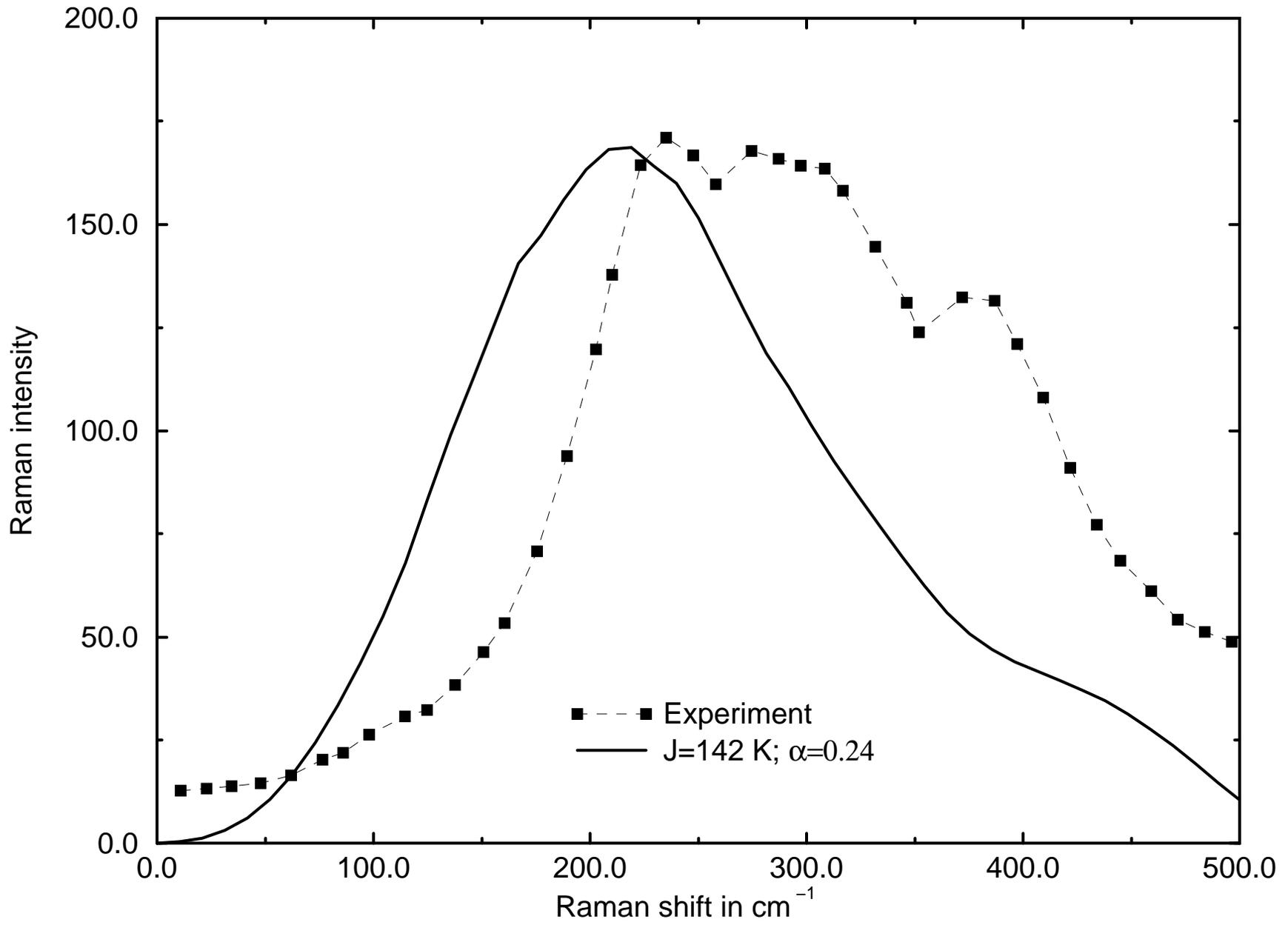

The results for the Raman intensity at $T = 20$ K with $J = 142$ K and
$\alpha = 0.24$ are shown in
Fig.\ \ref{raman} (solid line)
where they are compared with experimental results (squares)
obtained from a single crystal sample. Phonon lines at 184 cm$^{-1}$ and
330 cm$^{-1}$ have been subtracted from the experimental data. Since we
do not know the value of $A$ in our definition of the Raman operator $R$,
we have normalized the maximal theoretical value of the intensity to the
maximal experimental value. A uniform background of 50 counts has also
been subtracted from the experimental data. We see that there is a
reasonable agreement between theory and experiment as far as the
observed continuum is concerned. The shoulder observed experimentally at
$\sim$ 390 cm$^{-1}$ is not understood and our calculations show no
indications of this feature. We also see that 
the maximum of the experimental data
is shifted slightly toward larger energies than the theoretical results.
This could either be due to interchain couplings and phonon effects that
have been neglected or a mismatch in the parameters we have chosen. 
Our calculations are in good agreement with results
from exact diagonalization \cite{rc} as well as those 
of Singh and collaborators \cite{Singh_96} who have studied 
the same problem recently using a finite temperature Lanczos method.
We have also calculated the Raman intensities at higher temperatures. We find
that the continuum broadens out as temperature is increased and there is
no accumulation of spectral weight at lower energies as seen in
experiment \cite{Raman}. 

From the results shown in Fig.\ \ref{raman}, we conclude that the 
Hamiltonian (\ref{ham}) provides a good description of the observed Raman
continuum owing solely to the presence of the {n.n.n.} interaction.
These results as well as  the reasoning behind them also lead us to conclude
that there would be no inelastic Raman intensity in one
dimensional chains such as KCuF$_3$ where {n.n.n.} interactions are
negligible. In such materials, while INS can observe the {\it
one-magnon} continuum, the {\it
two-magnon} continuum cannot be observed in Raman scattering. 
These conclusions are indeed borne out by experiments \cite{KCuF}.
\section{Summary}

To summarize the results of this study, we have proposed a mean field
solution to the antiferromagnetic Heisenberg chain with {n.n.} and n.n.n.
interactions. This was motivated by recent interest evinced in the
spin-Peierls compound CuGeO$_3$ where it is suspected that n.n.n.
interactions play an important role. Our mean field solution provides a
self consistent way to include the effects of frustration and calculate
physical quantities of interest as a function of temperature.
Though we have limited our attention to experimental results in
CuGeO$_3$, the results in this paper can be used to study the effects of
frustration in generic quasi 1D systems.

As a first application, we have evaluated the temperature
dependent spin-wave velocity, $v_s(T)$ and found that (i)
$v_s(T)$ decreases rapidly for $\alpha=0$ with increasing $T$, 
$\alpha$ being the ratio between the {n.n.n.} and {n.n.} interactions and 
(ii) this decrease is much less pronounced for $\alpha\ne0$.
We propose that this fact could be exploited to determine
the frustration parameter $\alpha$
directly from measurements of $v_s(T)$ 
{\it e.g.}, by inelastic neutron scattering studies 
in other quasi 1D systems.

We then used the mean-field theory  to evaluate
the static susceptibility $\chi$.
A comparison with experimental results shows
that the inclusion of frustration does reproduce the suppression of
susceptibility at finite temperatures seen experimentally. 
We have been able to identify the physical reason for the
suppression of $\chi$ as a function of $\alpha$ 
with the differing behavior of the respective $v_s(T)$.
We have also presented the results of
a spin wave calculation for an anisotropic two dimensional
Heisenberg model which do not, on the other hand, 
compare well with the
experimental results for the susceptibility. This suggests that a one
dimensional model with frustration is a more appropriate starting point
to model the homogeneous phase of CuGeO$_3$. 

We also used the mean field solution to calculate the two-magnon Raman
intensity from a one dimensional antiferromagnetic chain and
showed that frustration is necessary for obtaining a
non-vanishing spectral weight.
Considering the fact that {n.n.n.} interactions are needed to
produce a good fit to the static susceptibility of CuGeO$_3$, using the
same idea to explain the Raman continuum provides a consistent
picture as far as modeling goes. Quite apart from the relevance to
CuGeO$_3$, this method of computing Raman intensity can be used to study
experimental results now available in Sr$_2$CuO$_3$ and SrCuO$_2$
\cite{srcuo}.
Clearly, a complete understanding of CuGeO$_3$ calls for the
inclusion of interchain and phonon effects. The
mean field solution presented in this paper provides a way to
incorporate these effects systematically. Work is in progress on these issues
and results will be reported in a forthcoming publication. 

\acknowledgements
This work was supported by the Deutsche 
For\-schungs\-gemein\-schaft, 
the Graduierten\-kolleg ``Fest\-k\"{o}rper\-spekt\-roskopie''.
V.~N.~M. acknowledges useful correspondence with Guillermo Castilla.

\begin{figure}
\caption{Spinon velocity as a function of temperature $T$,
obtained by the solitonic mean-field theory.
The long-dashed line shows results for 
$J = 115$ K and $\alpha = 0$. The solid line is obtained with
$J = 142$ K and $\alpha = 0.24$.  The values of $J$ are chosen such that
the velocities at zero temperature are the same in both cases.
\label{vel}}
\end{figure}

\begin{figure}
\caption{Experimental susceptibility of CuGeO$_3$ 
(filled squares, \protect\cite{Weiden_96}) as a function of temperature
compared with results from the mean field theory of the $J_1$ - $J_2$
model. The long-dashed line shows results with $J = 88$ K and $\alpha =
0$. The dotted line shows results with $J = 115$ K and $\alpha = 0$. 
The solid line is obtained with $J = 142$ K and $\alpha = 0.24$.
\label{susc_1D}}
\end{figure}

\begin{figure}
\caption{A comparison between the experimental 
susceptibility of CuGeO$_3$ 
(filled squares, \protect\cite{Weiden_96}) and results
obtained from a 2D spin wave theory with $J_c = 115$ K and $J_b = 0.1J_c$.
\label{susc_2D}}
\end{figure}

\end{document}